\documentclass[aps,prl,twocolumn,showpacs]{revtex4}
\usepackage{amssymb}

\usepackage{graphicx}

\begin{document}

\title{Proximity effect and its enhancement by ferromagnetism in
high-temperature superconductor-ferromagnet structures. }
\author{A.F. Volkov$^{1,2}$, K. B. Efetov$^{3}$.}

\address{Theoretische Physik III,\\
Ruhr-Universit\"{a}t Bochum, D-44780 Bochum, Germany\\
$^{(2)}$Institute for Radioengineering and Electronics of Russian Academy of\\
Sciences,11-7 Mokhovaya str., Moscow 125009, Russia\\
$^{(3)}$L.~D.~Landau Institute for Theoretical Physics RAS, 119334
Moscow, Russia}

\begin{abstract}
We consider a bi-layer consisting of a $d-$wave layerd superconductor and
diffusive ferromagnet with a domain wall (DW). The $c-$axis in the
superconductor and DW in the ferromagnet are assumed to be perpendicular to
the interface. We demonstrate that in such a heterostructure the
inhomogeneous exchange field enhances the proximity effect. It is shown
that, whereas in the absence of the exchange field the $d-$wave condensate
decays in the normal metal on the mean free path $l$, the superconductivity
penetrates the ferromagnet along the DW over much larger distances. This
happens because the presence of DW results in a generation of an odd
frequency triplet $s$-wave component of the condensate. The phenomenon
discovered here may help to explain a recent experiment on high temperature
superconductor-ferromagnet bi-lyers.
\end{abstract}

\maketitle

Proximity effect (PE) in superconductor/non-superconductor heterostructures
remains a popular topic of both experimental and theoretical study for
several decades. It is well known that the superconducting condensate can
penetrate a normal metal (N) over long distances $\sim $ $\xi _{N}=\sqrt{%
D/2\pi T},$ where $D$ is the diffusion coefficient in N and $T$ is
temperature \cite{deGennes}. In the limit of low temperatures $T\rightarrow
0 $ the length $\xi _{N}$ can be very large and this is a long range PE.

The situation is different in superconductor-ferromagnet (S/F)
heterostructures. If the exchange field $h$ in the ferromagnet is
homogeneous, the Cooper pairs penetrate a diffusive ferromagnet over
considerably shorter distances $\xi _{h}=\sqrt{D/h}$. Usually $T\ll h$ so
that the length $\xi _{h}$ is much shorter than $\xi _{N}$ \cite%
{BuzdinRMP,BVErmp} and PE is considerably suppressed with respect to the one
in the superconductor-normal metal (S/N) structures.

It turns out, however, that PE in S/F heterostructures can be long ranged
provided the exchange field is inhomogeneous. In this case a triplet odd
frequency $s$-wave component of the superconducting condensate is generated
and it can penetrate into the ferromagnet over a distance of the order $\xi
_{N}$, thus making PE similar to the one in the normal metals \cite%
{BVE01,BVErmp}. As a typical inhomogeneity one can think of either thick %
\cite{BVE01} or thin \cite{VE08} domain walls (DW) parallel (or
perpendicular \cite{VFE}) to the interface or spin-active interface \cite%
{Eschrig}. The prediction of the LRTC has been confirmed in several recent
experiments \cite{Keizer}.

The long range PE was usually studied in S/F heterostructures with
conventional superconductors ($s-$wave singlet superconductivity). At the
same time, S/F heterostructures can be fabricated using high temperature
superconductors (HTS) \cite{Pena,Millo06}, and study of PE in such systems
is also important and interesting.

In a recent work, the authors of Ref.~\cite{Millo06} using the scanning
tunnelling spectroscopy measured the density of states (DOS) $\nu (\epsilon
) $ at the outer surface of the ferromagnet in a bi-layer YBCO/F (a magnetic
material SrRuO was used as the ferromagnet). They found a dip in $\nu
(\epsilon )$ at energies $\epsilon =eV\lesssim 10meV$ in bi-layers with the
thickness of the ferromagnet $\lesssim 26nm$, whereas the \textquotedblleft
magnetic\textquotedblright\ length $\xi _{h}$ was estimated to be $\sim 3nm,$
that is, much shorter than the thickness of the F layer. Another interesting
finding was that the change of DOS occured only in the vicinity of DW. As a
possible reason for the observed effects, the authors of Ref.~\cite{Millo06}
considered LRTC. However, believing that this component had to spread all
over the sample rather than to be located near DW, they finally ruled out
this possibility.

Independent of whether LRTC was observed in Ref. \cite{Millo06} or not, it
is clear that theory of the HTS/F heterostructures is in a great demand.
Although the superconducting pairing in both conventional superconductors
and HTS cuprates is singlet, one cannot just use for HTS the theory
developed previously for the conventional superconductors because of a
special symmetry of the order parameter $\Delta $ in these materials.

The HTS cuprates are layered compounds with a weak coupling between the
layers. The symmetry of the order parameter in HTS differs from that in the
BSC superconductors \cite{vanHarlingen}. In the simplest version the order
parameter $\Delta $ in HTS has the structure: $\Delta (p)=\Delta
_{0}(p_{a}^{2}-p_{b}^{2})/p^{2}=\Delta _{0}\cos 2(\phi -\phi _{0})$, where $%
\phi $ is the azymuthal angle in the $(a,b)$ plane normal to the $c$
crystallographic axis, $p_{a}$ and $p_{b}$ are projections of the momentum
on the axes $a$ and $b$. Such a dependence of $\Delta (p)$ in HTS leads to
many interesting phenomena specific to HTS. For example, the gap in the
excitation spectrum turns to zero for certain directions. In addition, the
sign-variable angular dependence of $\Delta $ is the reason for the sign
change of the Josephson current in tunnel junctions composed of two HTS with
different orientations of crystallographic axes \cite{vanHarlingen}. At the
same time, we are not aware about any discussion of peculiarities of PE in
HTS/F heterostructures distinguishing them from the PE in conventional S/F
systems.

In this Letter we consider proximity effects of the HTS in contact with
disordered ferromagnets containing domain walls. We concentrate on the most
interesting situation when the $c$-axis is perpendicular to the interface,
such that the layers of the HTS are parallel to the interface. This geometry
corresponds to the experiment \cite{Millo06} and is special because, in
cases when $c-$ axis is not orthogonal to the interface, PE is similar to
the one in an $s-$ wave S/N (or S/F) bi-layer \cite{Ng,Nazarov,Golubov}.

We will see below that in the geometry considered the superconducting $d-$%
wave condensate penetrates a normal metal with the mean free path $l$ over a
distance of order $l,$ which is much shorter than the length $\xi _{N}$,
characterizing the penetration of the $s-$wave superconductivity. One can
say that PE is almost absent in this case but, as we will show, it is
restored if one replaces the normal metal by a ferromagnet with domain walls
perpendicular to the interface. In this case an \textit{odd triplet }$s-$%
\textit{wave }component is induced near DW and this is just the LRTC
predicted in Ref. \cite{BVE01} for S/F heterostructures with conventional
superconductors. It penetrates over much longer distances of order $\xi _{N}$%
. Previously, inhomogeneities of the exchange field lead to generation of
the odd triplet $s-$wave condensate from the conventional $s-$wave singlet
superconductor. Now the presence of DW leads to formation of the same $s-$%
wave component but starting from the $d-$ wave singlet superconductivity.

Before we give details of our calculation, let us explain qualitatively why
this interesting effect occurs. First, part of the effect, namely the
destruction of the proximity with the $d$ $-$wave superconductor and normal
metal is quite simple. It is well known the impurities destroy the $d$-wave
pairing. In the case, when the $c$-axis is perpendicular to the interface,
the symmetry of the condensate penetrating into the normal metal is the same
as that in the superconductor and it is destroyed in completely the same way.

Adding a domain wall perpendicular to the interface brings a coordinate
dependence of the magnetic momentum and therefore leads to a distortion of
the momentum dependence of the superconducting condensate. One can say that
the symmetry of the condensate function in F is broken because all
quantities depend on the orientaion of the rosette in S with respect to the
plane of the DW. The new distorted rosette can be expanded in angular
harmonics but now there is no symmetry reason for vanishing the $s$-wave
component. Then, as previously for the case of the $s$-superconductors with
a DW wall, both the singlet and odd triplet component of this $s$-wave are
generated and the latter can penetrate the normal metal over long distances
along the DW.

Our theory is based on using the method of quasiclassical Green functions.
Within this approach one has to solve a proper equation \cite{Eilenberger}
for the Green functions and then calculate physical quantities.

In order to simplify the problem we assume a weak PE, which corresponds to a
small transmission through the HTS/F interface. In this case the amplitude
of the condensate function $f$ in the ferromagnet is small and the
Eilenberger equation can be linearized. As a result, this equation takes in
the ferromagnet the following form (see for example~\cite{BVErmp,VE08})

\begin{eqnarray}
&&\mathit{sgn}\omega \cdot l(\mathbf{n\cdot \nabla })\hat{\tau}_{3}\mathbf{%
\otimes }\check{f}\;+\kappa _{\omega }\check{f}\;-i\lambda _{h}\{\cos \alpha
(y)[\hat{\sigma}_{3},\check{f}]_{+}+  \nonumber \\
&+&\sin \alpha (y)\hat{\tau}_{3}\mathbf{\otimes }[\hat{\sigma}_{2},\check{f}%
]\}=\langle \check{f}(\mathbf{n,r})\rangle ,  \label{EilenbergerN}
\end{eqnarray}%
where $\kappa _{\omega }=1+2|\omega |\tau ,\lambda _{h}=h\tau \mathit{sgn}%
\omega ,\omega =\pi T(2n+1)$ and $\mathbf{n=p/}p$ is the unit vector
parallel to momentum. The products $h\tau ,|\omega |\tau $ are assumed to be
small. The angle brackets mean the angle average, $[\hat{\sigma}_{3},\check{f%
}]_{+}$ and $[\hat{\sigma}_{2},\check{f}]$ denote anticommutator and
commutator.

We consider the case when the magnetization is oriented along the $z-$axis
far away from the DW and rotates in the DW in the $\{y,z\}$ plane changing
asymptotically its sign (see Fig.~1). This rotation is described by the
angle $\alpha (y).$ The $c-$axis is assumed to be orthogonal to the
interface.

The quasiclassical Gor'kov function $\check{f}$ entering Eq. (\ref%
{EilenbergerN})\ is a $4\times 4$ matrix in spin ($\hat{\sigma}_{i}$) and
particle-hole ($\hat{\tau}_{i}$) spaces and it depends on coordinates $%
\mathbf{r=}\{x,y\}$ (here the subindex $i=0,x,y,z$ and $\hat{\sigma}_{0},%
\hat{\tau}_{0}$ are the unit matrices). This matrix is off-diagonal in the
particle-hole space and contains both the diagonal ($f_{\pm }$) and
off-diagonal elements in spin space. We use the boundary conditions derived
in Refs.~\cite{Zaitsev}

\begin{figure}[tbp]
\begin{center}
\includegraphics[width=2.3cm, height=3.5cm]{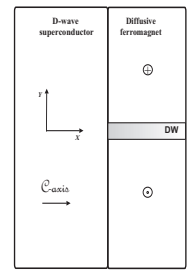}
\end{center}
\caption{Schematic picture of a HTS/F bi-layer with a domain wall DW of the
Neel type (the shadowed stripe). The signs in the circles denote the
direction of the magnetization in domains. }
\end{figure}

\begin{equation}
\mathit{sgn(}\omega \cdot n_{x})(\check{f}(n_{x},n_{y})-\check{f}%
(-n_{x},n_{y}))=\mathit{T}(n_{x})\hat{\tau}_{3}\mathbf{\otimes }\check{f}_{S}
\label{BCn}
\end{equation}%
where $\mathit{T}(n_{x})$ is the transmission coefficient, and $\check{f}%
_{S}=\hat{\sigma}_{3}\mathbf{\otimes }\hat{f}_{S},$ $\hat{f}_{S}=$ $\hat{\tau%
}_{2}f_{S}$ is the quasiclassical Gor'kov function in the HTS. This function
is assumed to be unperturbed by the weak PE and has the form: $f_{S}=\Delta (%
\mathbf{n})/\sqrt{\omega ^{2}+\Delta ^{2}(\mathbf{n})}$ with $\Delta (%
\mathbf{n})=\Delta _{0}(p_{a}^{2}-p_{b}^{2})/p^{2}$. Solving Eq. (\ref%
{EilenbergerN}) for $f$ with the boundary conditions, Eqs. (\ref{BCn}), one
can write the condensate contribution $\delta \nu (\epsilon )$ to the DOS at
the outer surface of the F film as $\delta \nu (\epsilon )=-(1/8)Tr(\hat{\tau%
}_{0}\mathbf{\otimes }\hat{\sigma}_{0}\langle \check{f}^{2}(\mathbf{n,}%
x)\rangle )|_{\omega =-i\epsilon ,x=d}$

The solution of Eq. (\ref{EilenbergerN}) can be found in a way similar to
the one used in Ref.~\cite{VE08}. First, we make a transformation
introducing a new matrix $\check{f}=\check{U}\check{f}_{U}\check{U}^{+},$
where the transformation matrix $\check{U}=\exp (i\hat{\tau}_{3}\hat{\sigma}%
_{1}\alpha (y)/2)$ describes a rotation in the spin space. Then, we
represent the matrix $\check{f}_{U}$ as a sum of symmetric and antisymmetric
in momentum space parts: $\check{f}_{U}=\check{s}+\check{a}.$ Substituting
the matrices $\check{s}$ and $\check{a}$ into Eq.~(\ref{EilenbergerN}), one
can express the antisymmetric matrix in terms of the symmetric one. For
example, for the elements $\hat{a}_{\pm }$ diagonal in the spin space we
obtain

\begin{equation}
\hat{a}_{\pm }\cdot \kappa _{h\pm }=-\mathit{sgn}\omega \cdot l\{(\mathbf{%
n\cdot \nabla })\hat{\tau}_{3}\mathbf{\otimes }\hat{s}_{\pm }+iQn_{y}\hat{s}%
_{1}\}  \label{AntisymN}
\end{equation}%
where $Q=\partial \alpha (y)/\partial y,\kappa _{h\pm }=\kappa _{\omega }\pm
2i\lambda _{h}$. It is clear from Eq. (\ref{AntisymN}) that the parity with
respect to the Matsubara frequency $\omega $ is different for the
antisymmetric part $\hat{a}_{\pm }$ that for the symmetric ones $\hat{s}%
_{\pm },\hat{s}_{1}.$

The off-diagonal component $\hat{a}_{1}$ describing LRTC can be found from a
similar equation obtained from Eq. (\ref{EilenbergerN}) by making the
replacement $\kappa _{h\pm }\rightarrow \kappa _{\omega }$ and $\hat{s}_{\pm
}\rightarrow \hat{s}_{1},$ $\hat{s}_{1}\rightarrow \hat{s}_{0}$ on the
right-hand side, where $\hat{s}_{0}=[\hat{s}_{+}+\hat{s}_{-}]/2$.

As concerns the diagonal $(\hat{s}_{\pm })$ and off-diagonal $(\hat{s}_{1})$
symmetric parts, they can be found from the equations

\begin{eqnarray}
&-&l^{2}(\mathbf{n\cdot \nabla })^{2}\hat{s}_{\pm }+\kappa _{h\pm }^{2}%
\hat{s}_{\pm }=\kappa _{h\pm }\langle \hat{s}_{\pm }\rangle +  \nonumber \\
&+in_{y}&[l^{2}(\mathbf{n\cdot \nabla })(Q\hat{\tau}_{3}\mathbf{\otimes }%
\hat{s}_{1})-lQ\kappa _{h\pm }\mathit{sgn}\omega \hat{a}_{1}],  \label{spmN}
\end{eqnarray}

\begin{eqnarray}
&-&l^{2}(\mathbf{n\cdot \nabla })^{2}\hat{s}_{1}+\kappa _{\omega }^{2}\hat{s}%
_{1}=\kappa _{\omega }\langle \hat{s}_{1}\rangle +  \nonumber \\
&+in_{y}&[l^{2}(\mathbf{n\cdot \nabla })(Q\hat{\tau}_{3}\mathbf{\otimes }%
\hat{s}_{0})-ilQ\kappa _{\omega }\mathit{sgn}\omega \hat{a}_{0}],
\label{s1N}
\end{eqnarray}%
where $\hat{a}_{0}=[\hat{a}_{+}+\hat{a}_{-}]/2$.

Using Eqs.~(\ref{BCn}) and (\ref{AntisymN}) one can write the boundary
conditions as

\begin{equation}
l|n_{x}|\frac{\partial \hat{s}_{\pm }}{\partial x}|_{x=0}=\mp \frac{\kappa
_{h\pm }}{2}\mathit{T}(n_{x})\hat{f}_{S},\text{ }\partial \hat{s}%
_{1}/\partial x|_{x=0,d}=0.  \label{BC1N}
\end{equation}

Eqs. (\ref{AntisymN}-\ref{BC1N}) completely describe the penetration of the
superconducting condensate for the geometry in Fig.1 that may correspond to
the experiment \cite{Millo06}. We solve these equations assuming that $lQ\ll
1$, which means that the mean free path $l$ is much smaller than the width
of the DW, $w$ ($Q\approx w^{-1}$). Then, in the main approximation we find
from Eqs.~(\ref{spmN}) and (\ref{BC1N}).

\begin{equation}
\hat{s}_{\pm }=\pm \frac{\mathit{T}(n_{x})}{2}\hat{f}_{S}\exp \left( -\frac{%
\kappa _{h\pm }x}{l|n_{x}|}\right) .  \label{SandAn}
\end{equation}%
The matrix $\hat{s}_{\pm }$ determines the amplitudes of the singlet and
triplet, $S_{z}=0,$ components. The angular average $\langle \hat{s}_{\pm
}\rangle $ is zero because $\langle \hat{f}_{S}\rangle =0.$ The amplitude of
the singlet component is proportional to $\hat{s}_{3}=(\hat{s}_{+}-\hat{s}%
_{-})/2$ and the amplitude of the triplet component with zero projection of
the total spin on the $z-$axis is proportional to $\hat{s}_{0}$. Both the
components decay over a distance of order $l$.

Obviously, this statement remains valid in the case of a HTS/N structure.
Putting $h=0$ in Eqs. (\ref{SandAn}) we see that only the components $\hat{s}%
_{3}$ and $\hat{a}_{3}$ are not equal to zero and that the condensate
penetrating the normal metal preserves the $d-$symmetry and decays on the
mean free path $l$, which is much shorter than the length $\xi _{N}$. Of
course, this is valid only when the axis $c$ is perpendicular to the
interface. Any deviation from this direction results in generation of the $%
s- $wave component penetrating the normal metal over the distance $\xi _{N}$ %
\cite{Nazarov,Golubov}.

Now we come to the demonstration that the presence of the ferromagnetism and
DW restores the long range penetration of the superconducting condensate.
This phenomenon occurs because such an inhomogeneous configuration of the
exchange field induces the odd triplet $s-$wave component of the
superconducting condensate with the projections of the spin $S_{z}=\pm 1$.

Substituting the matrices $\hat{s}_{0}$ and $\hat{a}_{0}$ into Eq.~(\ref{s1N}%
), one can obtain the equation for the triplet, $S_{z}=\pm 1,$ component, $%
\hat{s}_{1}$. We represent $\hat{s}_{1}$ in the form: $\hat{s}_{1}=\hat{s}%
_{1av}+\hat{s}_{1\sim }$ with $\hat{s}_{1av}\equiv \langle \hat{s}%
_{1}\rangle $, such that, $\hat{s}_{1av}$ does not depend on angles and $%
\langle \hat{s}_{1\sim }\rangle =0$. Equations describing $\hat{s}_{1av}$
and $\hat{s}_{1\sim }$ have quite different forms. Since $\langle \hat{s}%
_{1\sim }\rangle =0$, one can see from the left-hand side of the equation
for $\hat{s}_{1\sim }$ that the characteristic scale for decay of $\hat{s}%
_{1\sim }$ is the mean free path $l$ (in the considered dirty limit $\kappa
_{\omega }^{2}\approx 1$). In the equation for $\hat{s}_{1av}$ the second
term on the left hand side and the first term on the right hand side almost
compensate each other and one has: $\kappa _{\omega }^{2}\hat{s}%
_{1av}-\kappa _{\omega }\langle \hat{s}_{1}\rangle \approx (2|\omega |\tau )%
\hat{s}_{1av}.$ Therefore the angular averaged part of $\hat{s}_{1}$ decays
over a large distance of the order of $l/\sqrt{|\omega |\tau }$ $\sim \sqrt{%
D/T}$ and this is just the $S_{z}=\pm 1$ odd frequency triplet component.

In the main approximation the equation for $\hat{s}_{1av}$ reads

\begin{equation}
-l^{2}\frac{\partial ^{2}\hat{s}_{1av}(q)}{\partial x^{2}}+\kappa _{q}^{2}%
\hat{s}_{1av}(q)=3il^{2}(\partial Q/\partial y)_{q}\langle n_{y}^{2}\hat{s}%
_{0}(x)\rangle ,  \label{LRTCn}
\end{equation}%
$\hat{s}_{1av}(q)=\int dy\hat{s}_{1av}(y)\exp (iqy)$ and $\kappa
_{q}^{2}=6|\omega |\tau +q^{2}l^{2}$.

The function $\hat{s}_{0}(x)$ decays over a distance of the order $l,$
whereas the characteristic scale for $\hat{s}_{1av}(x,y)$ is much longer: $l/%
\sqrt{6\omega \tau }\approx \sqrt{D/2\pi T}$. Therefore, the term on the
right hand side can be replaced by $B\delta \left( x\right) ,$ where $B$
does not depend on $x.$ Solving Eq. (\ref{LRTCn}) in this approximation and
Fourier transforming back to $y,$ one can obtain the function $\hat{s}%
_{1av}(d,y)$ that determines the contribution $\delta \nu (\epsilon )$ of
the superconducting condensate to the density of states on the outer side of
the ferromagnet. The solution for the LRTC $\hat{s}_{1av}(x,y)$ at the $x=d$
can be presented in the form

\begin{equation}
\hat{s}_{1av}(d,y)=(3/4)i\hat{\tau}_{1}\lambda _{h}\cos (2\varphi _{0})%
\mathit{T}_{av}f_{Sav}S(y)  \label{s1av}
\end{equation}%
where $\mathit{T}_{av}$ and $f_{Sav}$ are the transmission coefficient and
condensate function in S averaged over angles with corresponding weights;
the Fourier component of the function $S(y)$ is equal to: $%
S(q)=l^{2}(\partial Q/\partial y)_{q}[\kappa _{q}\sinh (\kappa
_{q}d/l)]^{-1}.$

\begin{figure}[tbp]
\begin{center}
\includegraphics[width=3.5cm, height=4cm]{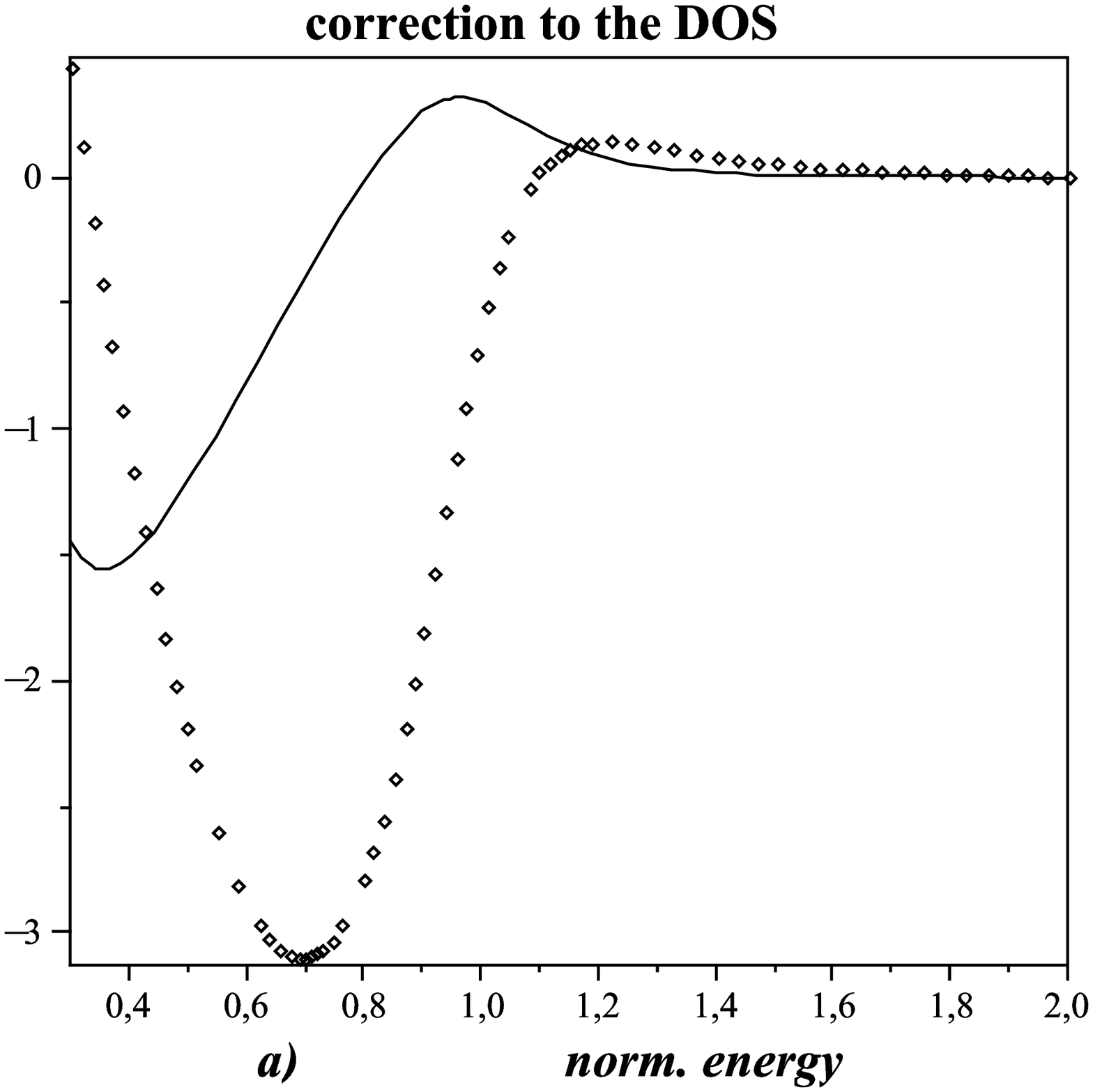} %
\includegraphics[width=3cm, height=4cm]{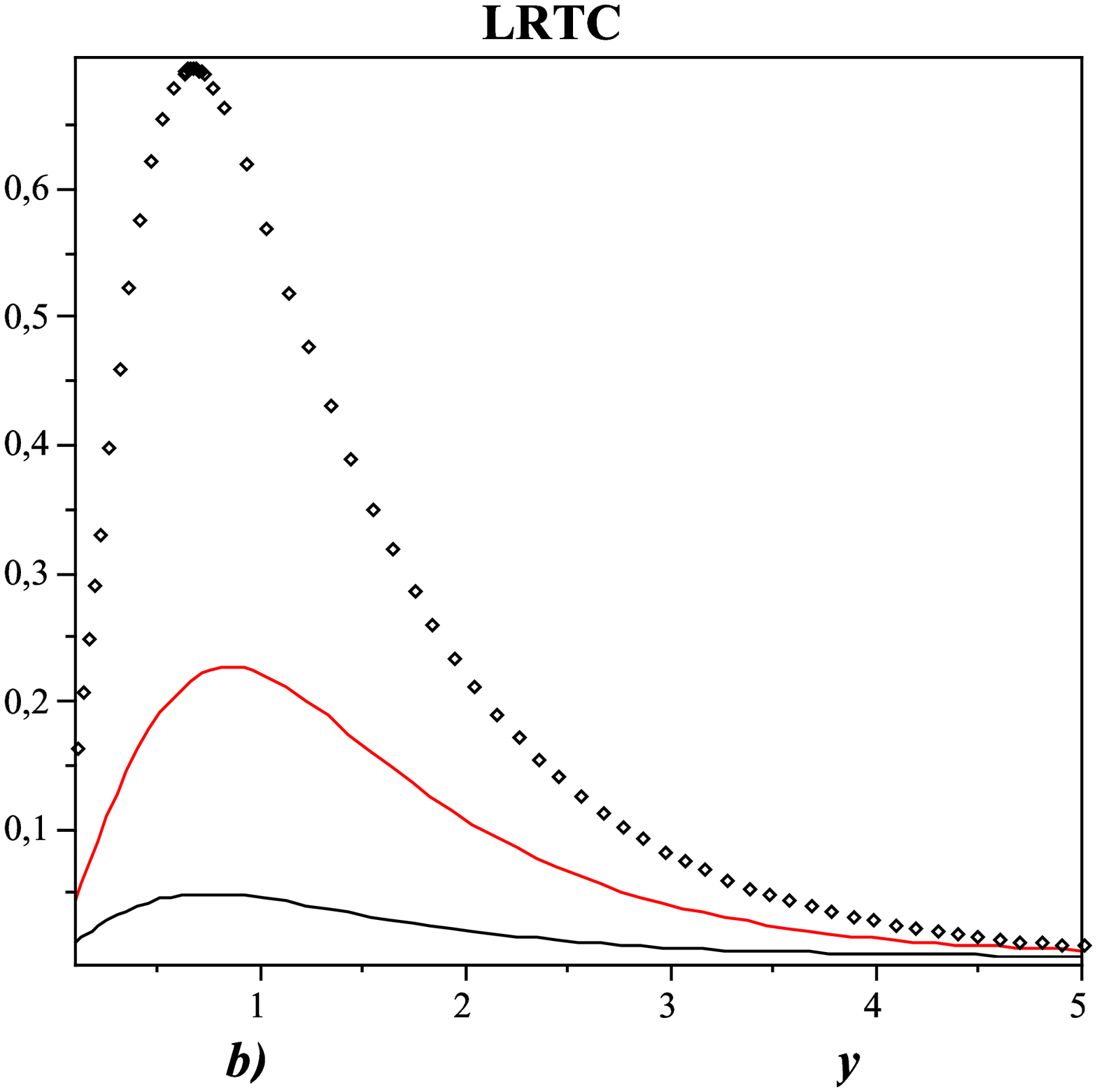}
\end{center}
\caption{a) Normalized correction to the DOS vs normalized energy $\protect%
\epsilon/\Delta_{0}$ for $d=2\protect\xi _{\Delta },m=\Delta_{0}\protect\tau%
_{m}=0.3$ (solid curve) and $d=\protect\xi _{\Delta },m=0.6$ (point curve),
where $\protect\xi _{\Delta }^{2}=D/(2\Delta_{0})$; b) Dependence of LRTC
amplitude on the normalized y-coordinate at $x=d$ for $w=\protect\xi _{%
\protect\omega }, d=0.5\protect\xi_{\protect\omega }$ (point curve); $w=%
\protect\xi _{\protect\omega },d=\protect\xi _{\protect\omega }$ (upper
solid curve) and $w=0.2\protect\xi _{\protect\omega }, d=\protect\xi _{%
\protect\omega }$ (lower solid curve). The y-coordinate is measured in units
$\protect\xi _{\protect\omega }$. }
\end{figure}

We emphasize that, in contrast to the singlet component $\hat{f}_{S}$, the
LRTC, $\hat{s}_{1av}$, is an odd function of the Matsubara frequency because
$\lambda _{h}\sim \mathit{sgn}\omega $ (see Eq.(\ref{EilenbergerN})). Thus,
the odd-frequency triplet $s-$wave component $\hat{s}_{1av}$ arises as a
result of the influence of the non-homogenous magnetization in the DW on the
$d-$wave singlet component. One can see that the LRTC is zero if the nodes
in the spectrum of the $d-$wave superconductor lie in the plane of the DW.

One can express the spatial dependence of this component in an explicit form
in limiting cases. If DW is broader than the LRTC penetration depth we can
write for the correction $\delta \nu (\epsilon )=\nu (\epsilon )-1$ to the
density of states: $\delta \nu _{eff}(\epsilon )=\delta \nu (\epsilon
)[(3/2)l^{2}\lambda _{h}\mathit{T}_{av}(\partial Q/\partial y)]^{-2}.$ We
calculated the function $\nu _{eff}\left( \epsilon \right) $ numerically and
represented in Fig. 2a an effective local correction to the DOS $\delta \nu
_{eff}(\epsilon )$ at $x=d$ caused by the LRTC.

In the opposite limiting case a simple form for the LRTC can be obtained if
the DW is approximated by a step-like function ($Q=\pi /w$ for $|y|<w/2$ and
$Q=0$ for $|y|>w/2)$. In this approximation we calculate the spatial
dependence of $S_{eff}(y)=-(w/l)S(y)$, for various values of $w$ and $d$
(Fig.2b). We see that the LRTC decays exponentially from DW over a long
distance of order $\xi _{N}.$

Note that spin-dependent scattering makes the characteristic length of the
LRTC decay shorter \cite{BVErmp,VE08}. The amplitude of LRTC contains a
small parameter $(l^{2}/wd).$ If this parameter is not small, the formulas
given above are valid qualitatively.

We believe that our results can help to explain the experimental data on the
density of states in a setup analogous to that considered here \cite{Millo06}%
, but more information about the DW, barrier transparency $\mathit{T}(n_{x})$
etc., is needed for a detailed comparison.

To conclude, we have demonstrated that the presence of a domain wall at the
interface of a high temperature, $d-$wave superconductor/ferromagnet
bi-layer leads to generation of an odd frequency triplet $s-$wave component
of the condensate. As a result, the superconducting condensate penetrates
the ferromagnet along DW over distances much exceeding the penetration depth
of the $d-$wave condensate into a normal metal. This explicitly demonstrates
enhancement of the proximity effect by an inhomogeneous exchange field.

We thank SFB 491 for financial support.

\bigskip


\bigskip

\begin{thebibliography}{99}
\bibitem{deGennes} P. G. de Gennes, Rev. Mod. Phys. \textbf{36}, 225 (1964).

\bibitem{BuzdinRMP} A. Buzdin, Rev. Mod. Phys. \textbf{77}, 935 (2005).

\bibitem{BVErmp} F.S. Bergeret, A.F. Volkov, K.B. Efetov, Rev. Mod. Phys.
\textbf{77}, 1321 (2005)

\bibitem{BVE01} F.S. Bergeret, A.F. Volkov, K.B. Efetov, Phys. Rev.Lett.
\textbf{78}, 4096 (2001).

\bibitem{Eschrig} M. Eschrig, J. Kopu, J.C. Cuevas, and G. Sch\"{o}n, Phys.
Rev. Lett. \textbf{90}, 137003 (2003).

\bibitem{VE08} A.F. Volkov and K.B. Efetov, Phys. Rev. B \textbf{78}, 024519
(2008).

\bibitem{VFE} A.F. Volkov, Ya.V. Fominov, and K.B. Efetov, Phys. Rev. B
\textbf{72}, 184504 (2005).

\bibitem{Keizer} R.S. Keizer et al., Nature \textbf{439}, 825 (2006); I.
Sosnin, et al., Phys. Rev. Lett. \textbf{96,} 157002 (2006).

\bibitem{vanHarlingen} D. J. Van Harlingen, Rev. Mod. Phys. \textbf{67}, 515
(1995); C.C. Tsuei, and J.R. Kirtley, \textit{ibid}. \textbf{72}, 969 (2000).

\bibitem{Pena} V. Pena et al., Phys. Rev. B \textbf{69}, 224502 (2004).

\bibitem{Millo06} I. Asulin, O. Yuli, G. Koren, and O. Millo, Phys. Rev. B
\textbf{74}, 092501 (2006); I. Asulin, O. Yuli, I. Felner, G. Koren, and O.
Millo, Phys. Rev. B \textbf{76}, 064507 (2007).

\bibitem{Ng} S. Kashivaya et al., Phys. Rev. B \textbf{60}, 3572 (1999);
Jian-Xin Zhu and C.S. Ting, \textit{ibid} \textbf{61}, 1456 (2000); Z.
Faraii and M. Zareyan, \textit{ibid} \textbf{69}, 014508 (2004); M. Freamat
and K. -W. Ng, \textit{ibid} \textbf{76}, 064507 (2007); N. Stefakanis and
R. Melin, J. Phys.: Condens. Matter \textbf{15}, 3401 (2003).

\bibitem{Nazarov} Y. Tanaka, Y. V.\ Nazarov, and S. Kashiwaya, Phys. Rev.
Lett. \textbf{90}, 167003 (2003); Y. Tanaka et al. Phys. Rev. B \textbf{69},
144519 (2004).

\bibitem{Golubov} Y. Tanaka and A. A. Golubov, Phys. Rev.Lett. \textbf{98},
037003 (2007); T. Yokoyama, Y. Tanaka, and A. A. Golubov, Phys. Rev. B
\textbf{73}, 094501 (2006).

\bibitem{Eilenberger} G. Eilenberger, Z. Phys. \textbf{214}, 195 (1968).

\bibitem{Zaitsev} A.V. Zaitsev, JETP 59, 1015 (1984).
\end{thebibliography}
\end{document}